\documentclass[10pt,conference,letterpaper]{IEEEtran}
\usepackage{times,amsmath,graphicx}
\usepackage{multirow,booktabs}
\usepackage{multicol}
\let\labelindent\relax
\usepackage{enumitem}
\usepackage{url}
\usepackage{hyperref}

\title{Coding of 3D Videos based on Visual Discomfort}
\author{%
{Dogancan Temel and Ghassan AlRegib}%
\vspace{1.6mm}\\
\fontsize{10}{10}\selectfont\itshape
School of Electrical and Computer Engineering, Georgia Institute of Technology\\
Atlanta, GA, 30332-0250 USA\\
\fontsize{9}{9}\selectfont\ttfamily\upshape
\{cantemel, alregib\}@gatech.edu
\vspace{1.2mm}\\
\fontsize{10}{10}\selectfont\rmfamily\itshape
}
\begin{document}

\onecolumn 

\begin{description}[labelindent=1cm,leftmargin=3cm,style=multiline]

\item[\textbf{Citation}]{D. Temel and G. AlRegib, "Coding of 3D videos based on visual discomfort," 2013 Asilomar Conference on Signals, Systems and Computers, Pacific Grove, CA, 2013, pp. 1356-1360.} \\

\item[\textbf{DOI}]{\url{https://doi.org/10.1109/ACSSC.2013.6810515}} \\

\item[\textbf{Review}]{Date added to IEEE Xplore: 8 May 2014} \\

\item[\textbf{Slides}]{\url{https://ghassanalregib.com/publications/}} \\

\item[\textbf{Bib}] {
@INPROCEEDINGS\{Temel2013\_Asilomar,\\ 
author=\{D. Temel and G. AlRegib\},\\ 
booktitle=\{2013 Asilomar Conference on Signals, Systems and Computers\},\\ 
title=\{Coding of 3D videos based on visual discomfort\},\\ 
year=\{2013\},\\ 
pages=\{1356-1360\},\\ 
doi=\{10.1109/ACSSC.2013.6810515\},\\ 
ISSN=\{1058-6393\},\\ 
month=\{Nov\},\}\\
} \\

\item[\textbf{Copyright}]{\textcopyright 2013 IEEE. Personal use of this material is permitted. Permission from IEEE must be obtained for all other uses, in any current or future media, including reprinting/republishing this material for advertising or promotional purposes,
creating new collective works, for resale or redistribution to servers or lists, or reuse of any copyrighted component
of this work in other works. }\\

\item[\textbf{Contact}]{\href{mailto:alregib@gatech.edu}{alregib@gatech.edu}~~~~~~~\url{https://ghassanalregib.com/}\\ \href{mailto:dcantemel@gmail.com}{dcantemel@gmail.com}~~~~~~~\url{http://cantemel.com/}}
\end{description} 

\thispagestyle{empty}
\newpage
\clearpage

\twocolumn

\maketitle

\begin{abstract}
We propose a rate-distortion optimization method for 3D videos based on visual discomfort estimation. We calculate visual discomfort in the encoded depth maps using two indexes: temporal outliers (TO) and spatial outliers (SO). These two indexes are used to measure the difference between the processed depth map and the ground truth depth map. These indexes implicitly depend on the amount of edge information within a frame and on the amount of motion between frames. Moreover, we fuse these indexes considering the temporal and spatial complexities of the content. We test the proposed method on a number of videos and compare the results with the default rate-distortion algorithms in the H.264/AVC codec. We evaluate rate-distortion algorithms by comparing achieved bit-rates, visual degradations in the depth sequences and the fidelity of the depth videos measured by SSIM and PSNR.
\end{abstract}

\hbox{}

 \begin{keywords}
 Rate distortion optimization, depth map, 3D video, perceptual quality, image fidelity, visual discomfort
 \end{keywords}
%


\section{Introduction}\label{sec:intro}

Free Viewpoint Video \emph{(FVV)} enables navigation inside a scene whereas Three-Dimensional Television \emph{(3DTV)} provides the depth perception to the end user. 3D view synthesis methods are either based on 2D depth maps or sparse 3D scene structures. In terms of video coding, depth map-based methods are more feasible and applicable because they do not require large bandwidth and the acquisition system can be as simple as a stereo camera. When 3D view synthesis methods are considered, Depth-Image-Based Rendering (\emph{DIBR}) is the most commonly preferred approach in the literature \cite{Fehn04}. In DIBR-based 3D view synthesis, a depth map is required for each color frame. Depth map is basically a grayscale image where each pixel value corresponds to the relative distance between the camera reference frame and world reference frame. By using the relative depth information in the depth map and camera setup parameters such as baseline, focal length, convergence distance and relative position, color pixels can be mapped to the world frame in 3D and then they can be projected to a new 2D camera frame (virtual view) at the receiver side. Users can have the 3D experience by feeding the stereo image pair that consists of reference and virtual view to the 3D display system.

We can encode the depth sequences using the Advanced Video Coding \emph{(AVC)} standard. A detailed overview of the \emph{H.264/MPEG-4 AVC} is provided in \cite{Puri2003}. As it is explained in \cite{Puri2003}, rate control approaches consist of three main steps: target bit allocation, virtual buffer based bit-rate control and adaptive quantization. In order to understand the structure of AVC and how it performs these main steps, authors in \cite{JVTNotes} explain the rate control in three levels as: GOP Level rate control, Picture Level rate control and Basic Unit Level rate control (optional).

In the GOP Level rate control, Quantization Parameter (\emph{QP}) is initialized based on the available channel bandwidth and QP for the rest of the pictures in the GOP are calculated according to the formulas described in the technical notes by  Joint Video Team \emph{(JVT)}  \cite{JVTNotes}. In the Picture Level rate control, control system consists of pre-encoding and post-encoding steps. We are interested in the pre-encoding stage of the stored pictures where controller performs Rate Distortion Optimization \emph{(RDO)} by setting  QPs of each picture. QP value assignment depends on the objective quality of the pictures and  Mean Absolute Difference (\emph{MAD}) is used in the MPEG-4 AVC standards. The last level of rate control is the Basic Unit Level rate control where rate control is performed for a group of continuous macro-blocks in addition to GOP and Picture Levels.

In this paper, we use Picture Level rate control for coding \emph{3D} videos. Since depth sequences are not directly presented to the end user, we need to use quality metrics that consider the depth perception of the Human Visual System \emph{(HVS)}. Authors in paper \cite{Kondoz2011} investigate the relationship between quality of synthesized view and the quality of the depth maps. They illustrate that PSNR and SSIM values of depth maps do not correlate well with the rendered view quality and they conclude that these metrics are not suitable for measuring degradation based on compression. In this paper, we compute the factors that can lead to visual discomfort instead of assessing the fidelity in the depth sequences. Objective quality assessment of 3-D videos based on visual discomfort was proposed in \cite{Solh2011} as \emph{3VQM}. Authors derive an ideal depth and define the error metric as the absolute value of the difference between the estimated depth and the ideal depth. Using the error metric, temporal and spatial characteristics of the DIBR-based 3-D videos are calculated in the form of  temporal outliers, temporal inconsistencies and spatial outliers. Effectiveness of \emph{3VQM} in capturing errors and inconsistencies is evaluated in \cite{Temel2013}. Authors validate \emph{3VQM} by showing that it is more accurate, coherent and consistent compared to PSNR and SSIM that are calculated over synthesized views.

Visual Discomfort Metric \emph{(VDM)} is introduced in this paper. It is basically a modified version of \emph{3VQM} to estimate perceived compression errors instead of depth map estimation errors. 3D videos are encoded with \emph{VDM} by analyzing the spatial and temporal characteristics of the depth sequences to make the video coding content-adaptive. The rest of this paper is organized as follows. Section \ref{sec:discomfortdefinition} describes visual discomfort estimation. In Section \ref{sec:ratedistopt}, the interaction between encoder and the visual discomfort estimation is explained. Distortion assessment is provided in Section \ref{sec:discomfortassesment} and rate-distortion analysis is discussed in Section \ref{sec:ratedisanalysis}. Finally, the concluding remarks are stated in Section \ref{sec:conc}.

\section{VISUAL DISCOMFORT ESTIMATION}\label{sec:discomfortdefinition}

Objective quality metrics are commonly used to asses video quality in streaming applications. These metrics are required to be real time and highly correlated with the subjective assessment. Pixel-based quality metrics such as MAD, MSE and PSNR are used because of the simplicity of implementation. However, these pixel-based methods do not correlate well with subjective tests. In the proposed work, we focus on visual discomfort instead of pixel-wise degradations.

Visual Discomfort Metric (\emph{VDM}) is an adaptive version of \emph{3VQM} \cite{Solh2011}. \emph{3VQM} was used to calculate depth map estimation errors. In contrast, in this paper, we assume that ground truth depth maps are error free. We estimate compression-based quality degradation by comparing the ground truth and processed depth maps that are used for rendering virtual views. The depth map error definition in \emph{3VQM} is given as:
   \begin{equation}\label{eq:dz1}
\Delta Z=|Z_{ideal} -Z_{GT}|,
    \end{equation}
where $Z_{ideal}$ is the ideal depth map that is defined in \emph{3VQM} as the depth map that results in distortion-free video and $Z_{GT}$ is the Ground Truth depth map. In \emph{VDM}, the error definition is modified by replacing ideal depth with ground truth depth and ground truth depth with compressed depth as follows:
    \begin{equation}\label{eq:dz2}
\Delta \hat{Z}=|Z_{GT} -Z_{processed}|,
    \end{equation}
where $\Delta \hat{Z}$ is the absolute value of the difference between the ground truth depth map and the processed depth map and $Z_{processed}$ is the depth map after compression. In the following parts, we define the visual discomfort indexes used in \emph{3VQM} and \emph{VDM}.

Compressing the depth map leads to spatial inconsistencies in the depth values. These discontinuities result in relocated pixels in the synthesized views, which cause visual discomfort. We measure spatial discomfort using the standard deviation of $\Delta \hat{Z}$ and call the quantity as Spatial Outlier \emph{(SO)}:
    \begin{equation}\label{eq:SO}
SO=STD(\Delta \hat{Z}).
    \end{equation}

Depth map error patterns can also temporally vary because of the compression artifacts. These artifacts result in impulsive intensity changes around textured regions and flickering around flat regions. Temporal variation of the depth map errors can be modeled by  the standard deviation of the difference between the depth map errors in consecutive frames. We define these errors as Temporal Outliers \emph{(TO)} as follows:
    \begin{equation}\label{eq:TO}
TO=STD(\Delta \hat{Z}_{t+1}-\Delta \hat{Z}_{t}),
    \end{equation}
where $STD$ is the standard deviation, $\Delta \hat{Z}_{t}$ and $\Delta \hat{Z}_{t+1} $ are the predicted depth map error in frame $t$ and $t+1$, respectively.
Temporal depth consistency is a significant factor in visual comfort. Temporal depth inconsistency can be measured by quantifying the excessive and fast changing disparities using the standard deviation of the difference of consecutive depth frames. We call this quantity as Temporal Inconsistencies \emph{(TI)} and is given as follows:
\begin{equation}\label{eq:TI}
TI=STD(\hat{Z}_{t+1}-\hat{Z}_{t}).
\end{equation}

%
%
%
%
%
%
%
%

In \emph{3VQM}, the pooling of these indexes is based on computing the complements of the indexes so that \emph{3VQM} decreases as the image gets distorted. The \emph{SO} index is masked with the logical intersection of \emph{SO} and \emph{TO} indexes to avoid considering visual discomfort sources more than once. Finally, the powers of the discomfort indexes in \emph{3VQM} are determined according to an offline training process and the combination is scaled with a coefficient. \emph{3VQM} formulation is given in \cite{Solh2011} as follows:
\begin{equation}\label{eq:3VQM}
3VQM=K(1-SO(SO \bigcap TO))^a(1-TO)^b(1-TI)^c ,
\end{equation}
where $K=5.0$, $a=8.0$, $b=8.0$, and  $c=6.0$.

In \emph{VDM}, \emph{TI} contributes to the discomfort metric as in \emph{3VQM}. However, power assignment of \emph{TO} and \emph{SO} are modified. Instead of calculating the powers of complement of the metrics, we calculate the power of discomfort metrics  and then take the complement. Moreover, power assignment of \emph{TO} and \emph{SO} are content-adaptive in \emph{VDM}. Spatial and temporal information indexes are calculated as defined in \cite{ITUTNotes}. To calculate spatial information index, luminance channel of each frame is filtered with a Sobel operator and then the standard deviation is computed over pixels. This procedure is performed for all frames and the maximum value represents the video sequence. Temporal information index is calculated by taking the difference between consecutive frames, calculating the standard deviation over the pixels for each frame and then selecting the maximum index over time. $S_{Inf}$ stands for the cube root of Spatial Information index and $T_{Inf}$ stands for the cube root of Temporal Information index as shown in Equation (\ref{eq:SInf}) and Equation (\ref{eq:TInf}), respectively. We use $S_{Inf}$ as the power of \emph{SO} index and $T_{Inf}$ as the power of \emph{TO} index as in Equation (\ref{eq:VDM}).
    \begin{equation}\label{eq:SInf}
    S_{Inf}=\sqrt[3]{max_{time}\{std_{space}[Sobel(F_n)] \}},
    \end{equation}
  \vspace{-0.2cm}
    \begin{equation}\label{eq:TInf}
    T_{Inf}=\sqrt[3]{max_{time}\{std_{space}[F_n(i,j)-F_{n-1}(i,j)] \}},
    \end{equation}
where $F_n$: current frames, $F_{n-1}$: previous frame, $std$: standard deviation, $max_{time}$: max operator that selects the maximum index over time (over all the frames in the video).
\vspace{0.2cm}

The \emph{TI} index can capture the depth map estimation errors as described in \emph{3VQM} \cite{Solh2011}. However, when we asses the change in the perceived quality with respect to varying compression errors, \emph{TI} is not highly correlated with the subjective results. As depth videos are quantized more coarsely, depth maps become smoother. Difference of consecutive frames becomes less significant when depth maps are smoothed, which leads to a lower \emph{TI} index. Formulation of  \emph{VDM} is given in Equation (\ref{eq:VDM}). When we assign a negative value to parameter \emph{c}, \emph{TI}  will decrease the value of \emph{VDM} as we quantize the depth videos more coarsely. However, negative powers of the complement of \emph{TI} linearly decreases with the increasing quantization parameter which is correlated with PSNR more than SSIM or perceived quality. Therefore, we exclude \emph{TI} from \emph{VDM} by assigning $0.0$ to parameter \emph{c} in Equation (\ref{eq:VDM}). Finally, we take the logical intersection of \emph{SO} and \emph{TO} indexes out of the equation since visual discomfort becomes more severe when we have both type of distortions at the same pixel locations. The resulting measure is given as follows:
    \begin{equation}\label{eq:VDM}
VDM=K(1-SO^a)(1-TO^b)(1-TI)^c ,
    \end{equation}
where $K=1.0$, $a=S_{Inf}$, $b=T_{Inf}$ and  $c=0.0$.

\begin{figure*}[htbp!]
\centering
\includegraphics[width=0.75\linewidth]{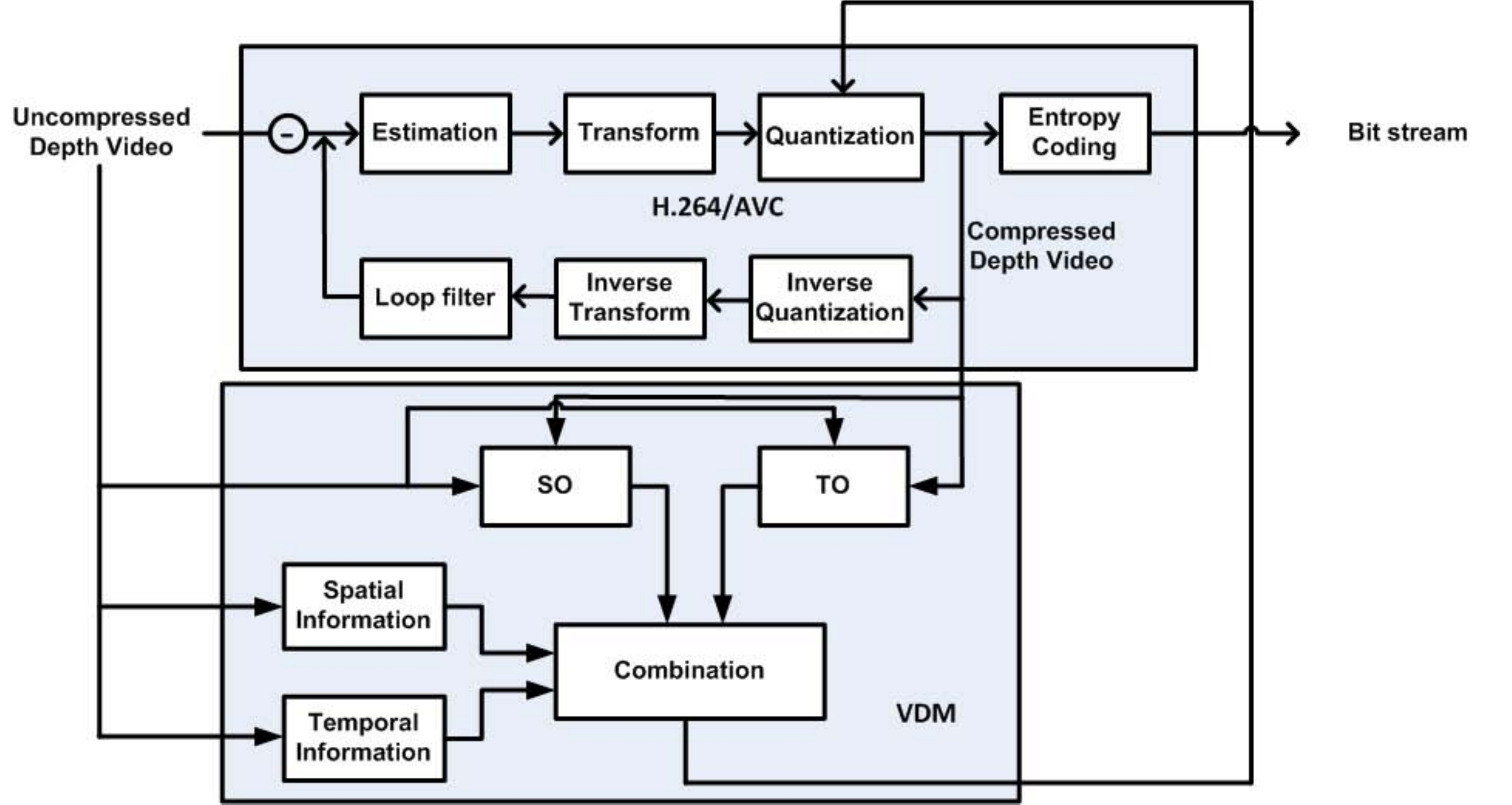}
\caption{Rate Distortion Optimization Pipeline}
\label{fig:RDpipeline}
\vspace{-0.6cm}
\end{figure*}

\section{RATE-DISTORTION OPTIMIZATION}\label{sec:ratedistopt}

The H.264/AVC pipeline is shown in Figure \ref{fig:RDpipeline} \cite{AVCmodel}. At the AVC encoder, we have an access to both original and processed (compressed) videos. Therefore, full reference metrics can be used to measure distortion. In the default rate control mechanism of H.264, Mean Absolute Difference (\emph{MAD}) is used as the distortion metric. However, we need to use distortion metrics that consider the structure of the content and perception instead of basic pixel-wise comparisons. Especially when we encode depth sequences instead of color sequences, distortion metrics should correlate with the errors in rendered 3D views. As described in Section \ref{sec:discomfortdefinition}, we use \emph{VDM} to estimate the depth map compression errors. Frames are compressed with the maximum quantization parameter (\emph{QP}) that satisfies minimum \emph{VDM} requirements which is calculated as the mean \emph{VDM} of all the frames when they are quantized with constant QP in the range of $30$ to $49$. At first we encode depth sequences with constant QPs and obtain a lookup table for minimum \emph{VDM} values for each sequence. Then, we initialize the AVC encoder and compress the depth frames. \emph{VDM} is calculated by using the ground truth frame and the compressed frame, if \emph{VDM} is higher than the threshold we increase the QP for the next frame, otherwise we decrease the QP. We set minimum QP as $30$ and maximum QP as $50$.

\vspace{0.10in}


\section{DISTORTION ASSESSMENT }\label{sec:discomfortassesment}
\begin{figure} [ht!]
\begin{minipage}[b]{0.48\linewidth}
  \centering
\includegraphics[width=\linewidth]{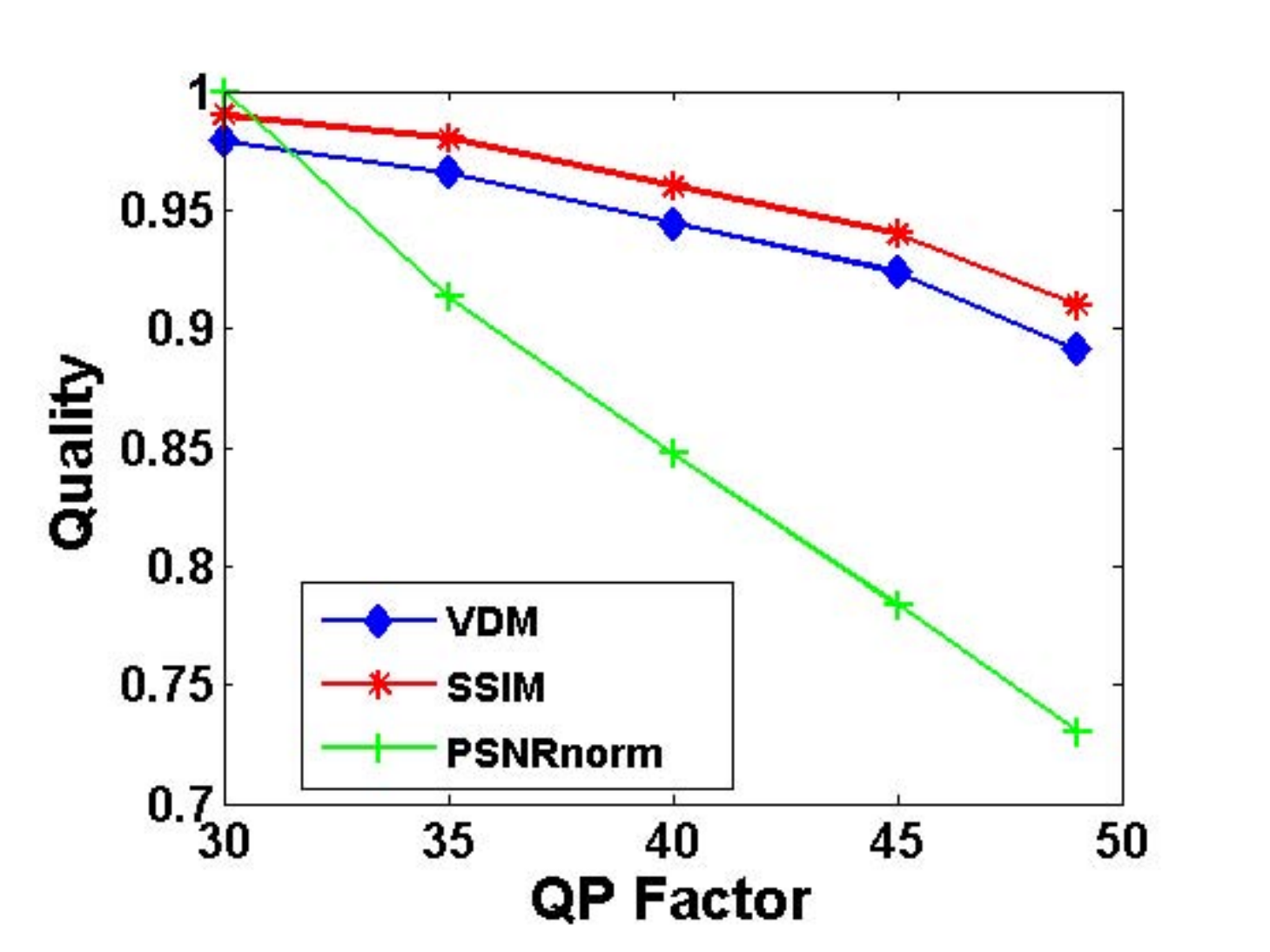}
  \centerline{\footnotesize{(a) Balloons Depth}  }
\end{minipage}
\begin{minipage}[b]{0.48\linewidth}
  \centering
\includegraphics[width=\linewidth]{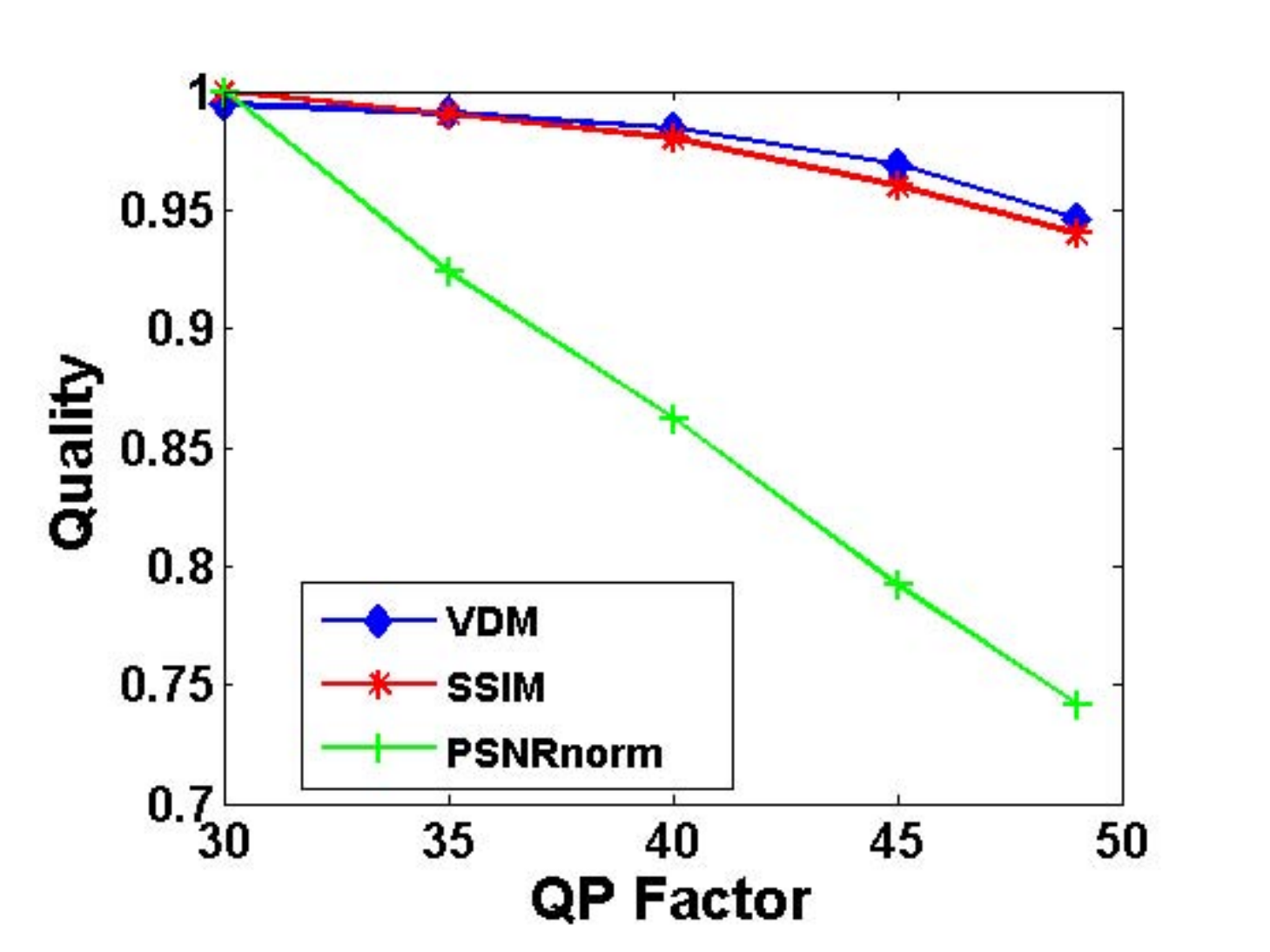}
  \centerline{\footnotesize{(b) Champagne Tower Depth  }}
\end{minipage}
 \begin{minipage}[b]{0.48\linewidth}
  \centering
\includegraphics[width=\linewidth]{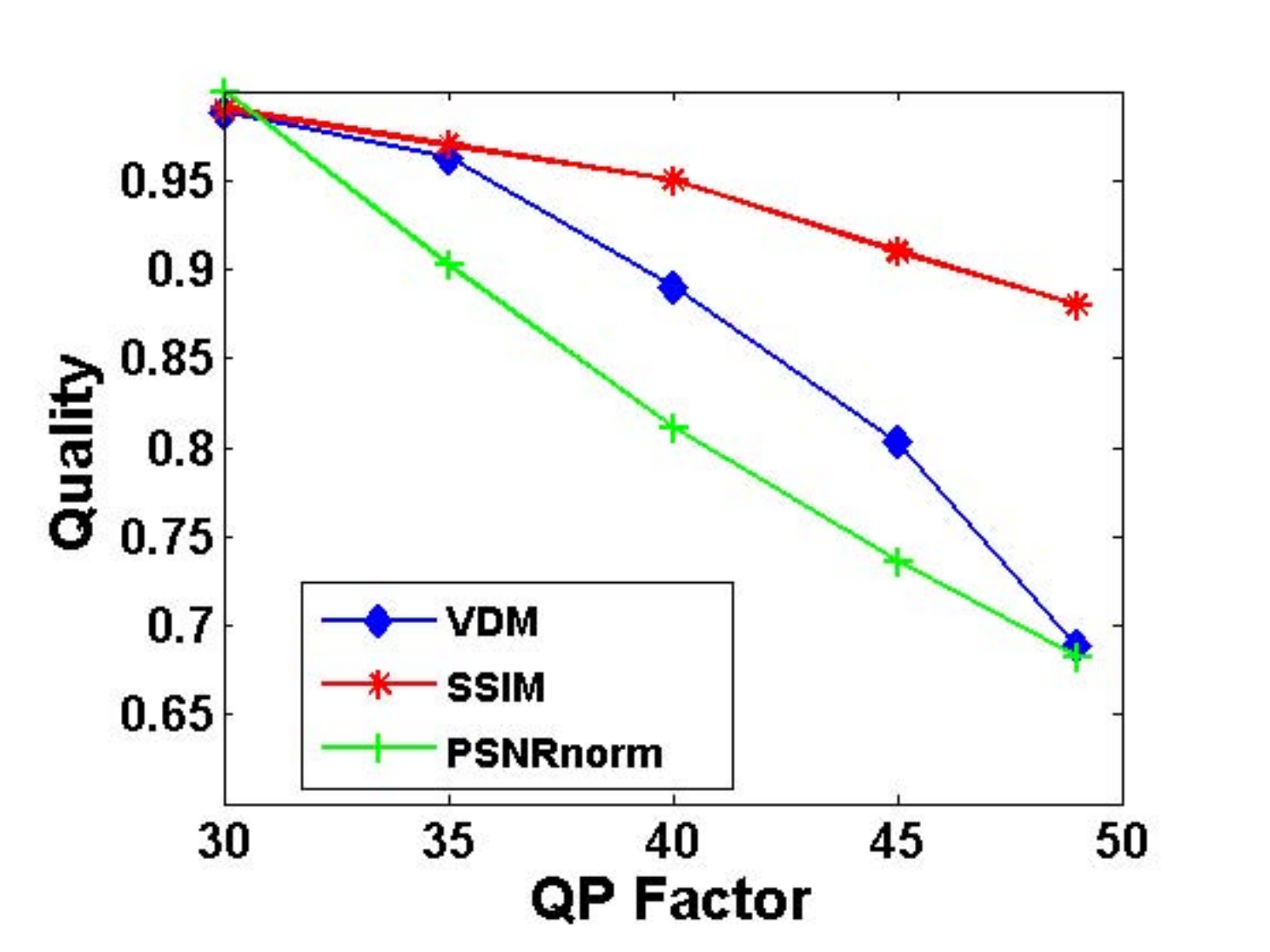}
  \centerline{\footnotesize{(c) Kendo Depth  }}
\end{minipage}
 \begin{minipage}[b]{0.48\linewidth}
  \centering
\includegraphics[width=\linewidth]{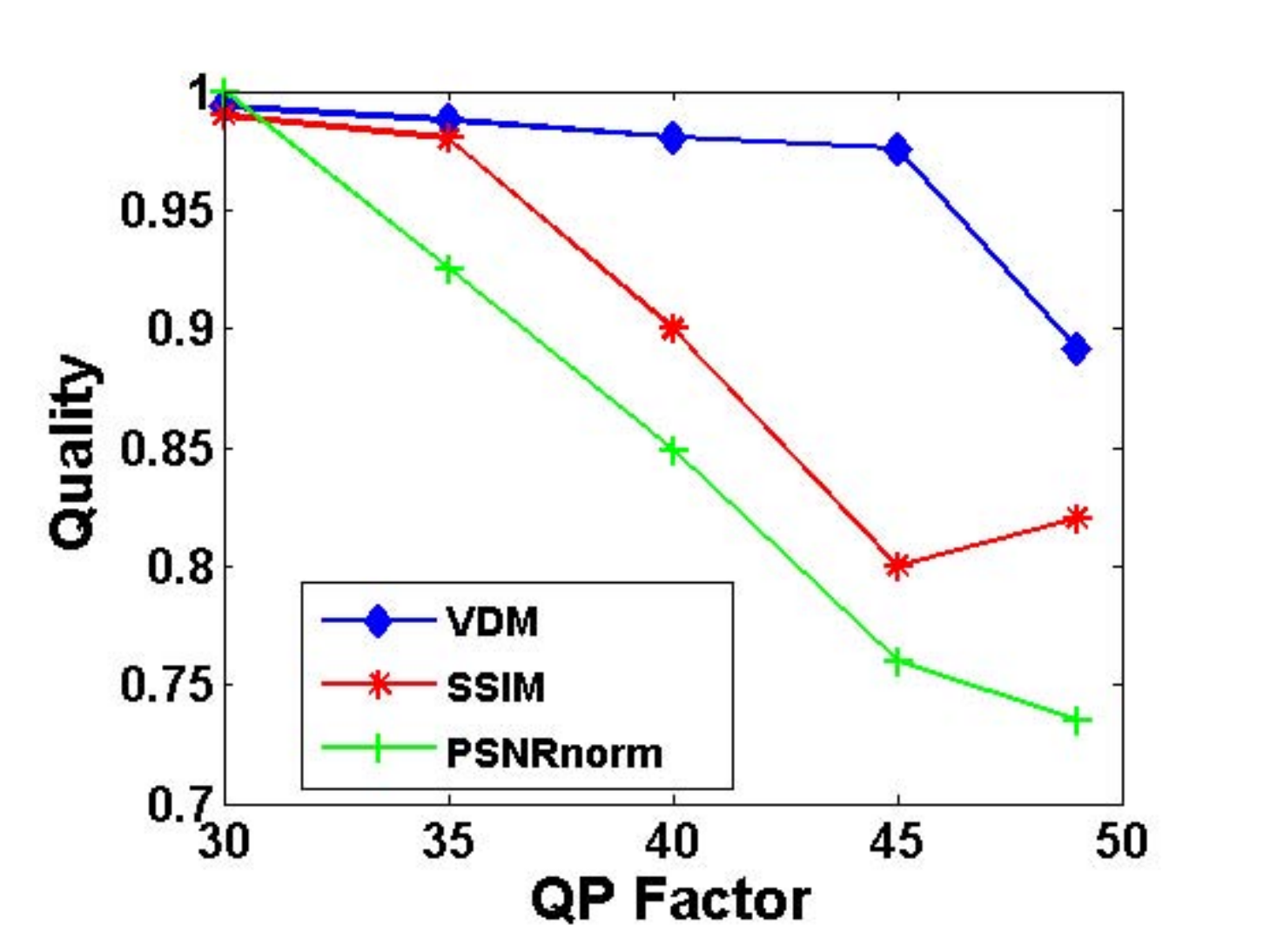}
  \centerline{\footnotesize{(d) Lovebirds Depth  }}
\end{minipage}
 \begin{minipage}[b]{0.48\linewidth}
  \centering
\includegraphics[width=\linewidth]{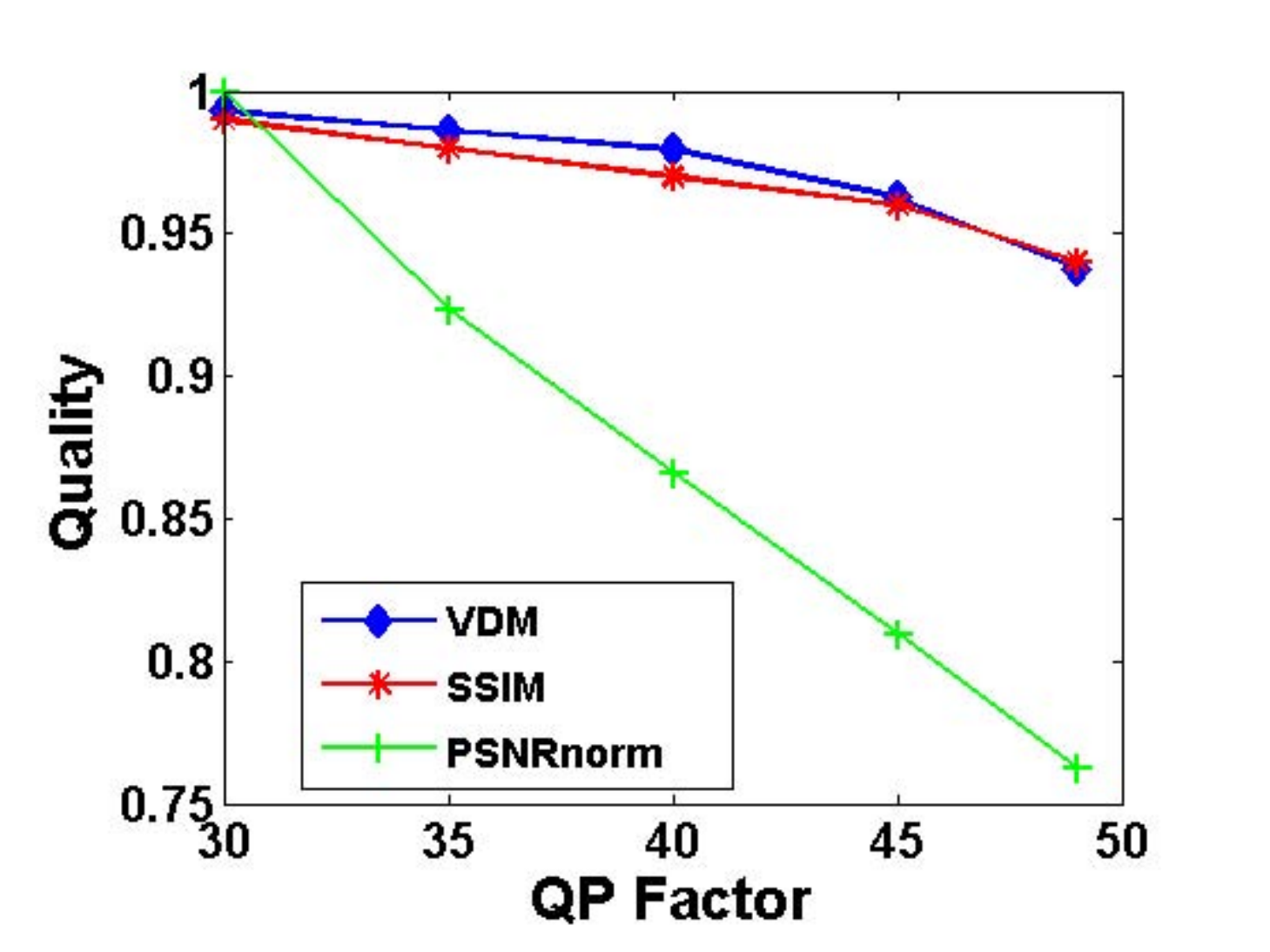}
  \centerline{\footnotesize{(e) Pantomime Depth  }}
\end{minipage}

\caption{Objective quality results for compressed (lossy) depth video sequences}
\label{fig:qualitymetrics}
\vspace{-0.2in}
\end{figure}

Lossy compression methods lead to artifacts, which result in visual distortions. These distortions distract the users and degrade the quality of experience. We perform compression using ver. 18.5 of H.264/AVC reference software and CABAC is used as the entropy coding method. The video sequences used in this work are obtained from \emph{3DMobile} project video database and they can be sorted as follows: \texttt{Balloons}, \texttt{Champagne Tower}, \texttt{Kendo}, \texttt{Lovebirds} and \texttt{Pantomime} \cite{3DMobile}. In this section, we show how \emph{VDM}, SSIM and PSNR perform under varying compression ratios. \emph {QP} is set to \emph{30}, \emph{35}, \emph{40}, \emph{45} and \emph{49}. Quality metrics are calculated using ground truth and processed depth sequences. In order to represent the metrics at the same plot, PSNR is normalized as shown in Figure \ref{fig:qualitymetrics}.

\begin{figure*} [ht!]

\begin{minipage}[b]{0.19\linewidth}
  \centering
\includegraphics[width=0.9\linewidth]{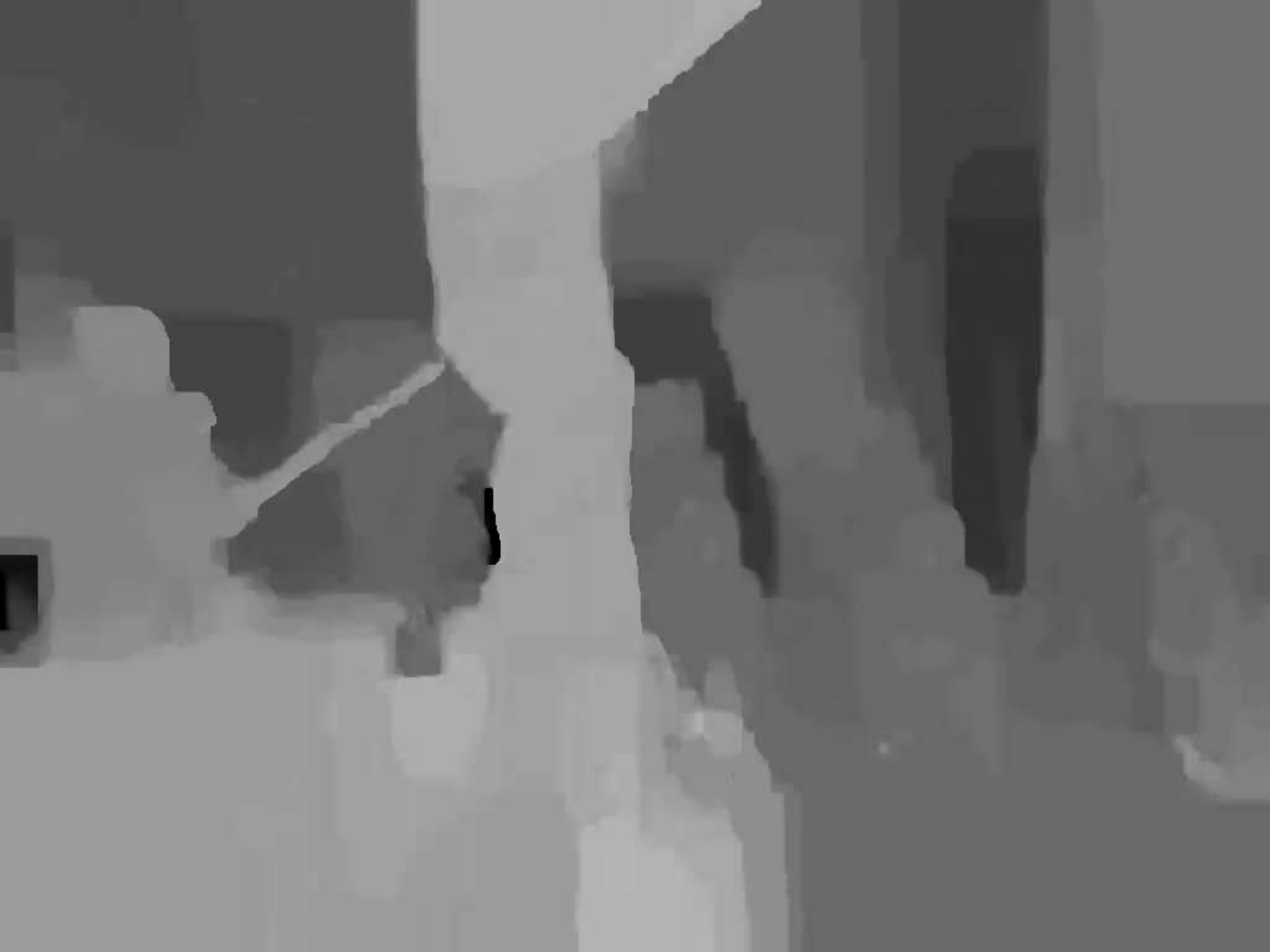}
  \centerline{\footnotesize{(a) Kendo Depth QP=30  }}
\end{minipage}
 \hfill
 \begin{minipage}[b]{0.19\linewidth}
  \centering
\includegraphics[width=0.9\linewidth]{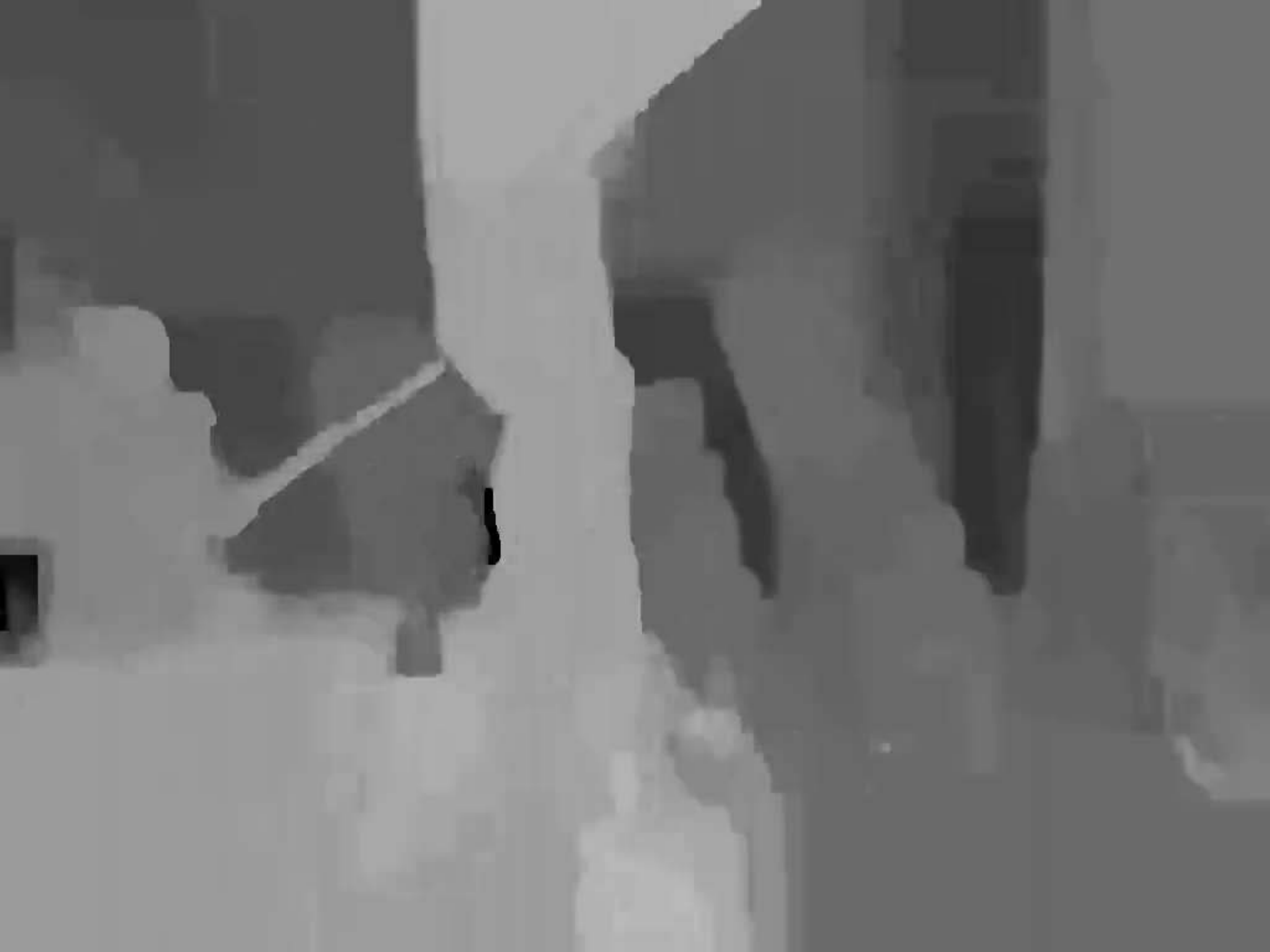}
  \centerline{\footnotesize{(b) Kendo Depth QP=35  }}
\end{minipage}
 \hfill
 \begin{minipage}[b]{0.19\linewidth}
  \centering
\includegraphics[width=0.9\linewidth]{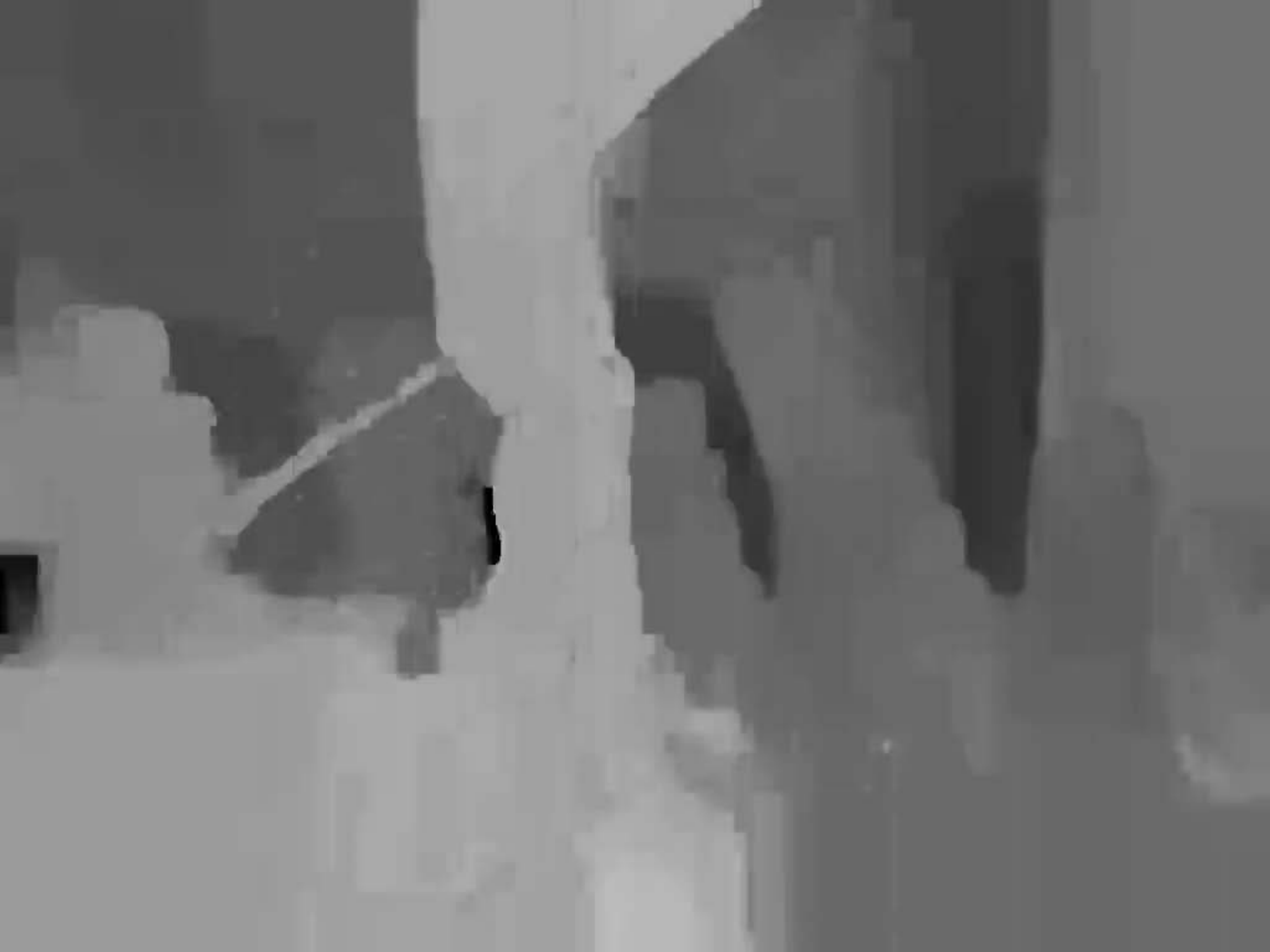}
  \centerline{\footnotesize{(c) Kendo Depth QP=40  }}
 \end{minipage}
 \hfill
 \begin{minipage}[b]{0.19\linewidth}
  \centering
\includegraphics[width=0.9\linewidth]{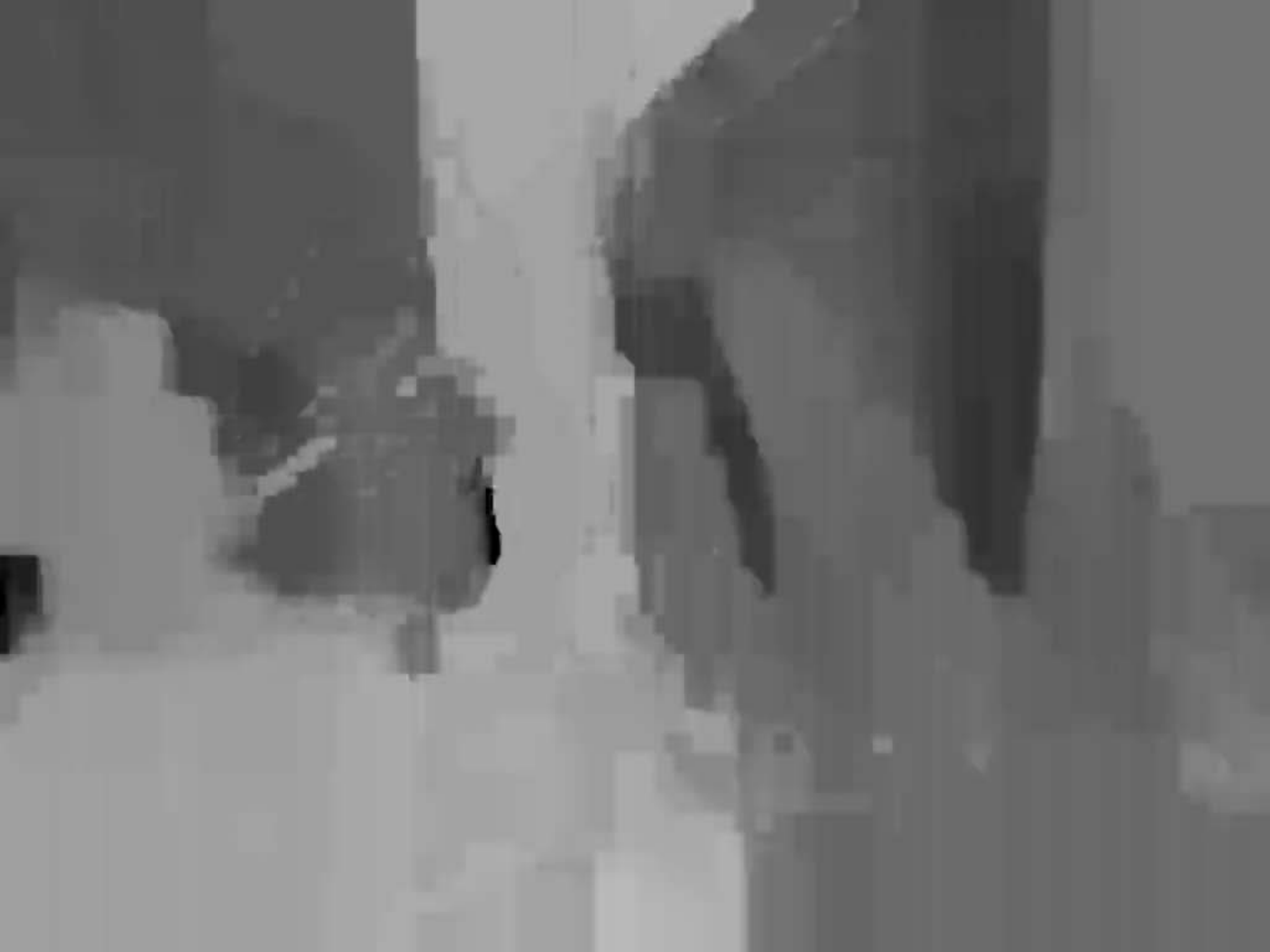}
  \centerline{\footnotesize{(d) Kendo Depth QP=45  }}
\end{minipage}
\hfill
 \begin{minipage}[b]{0.19\linewidth}
  \centering
\includegraphics[width=0.9\linewidth]{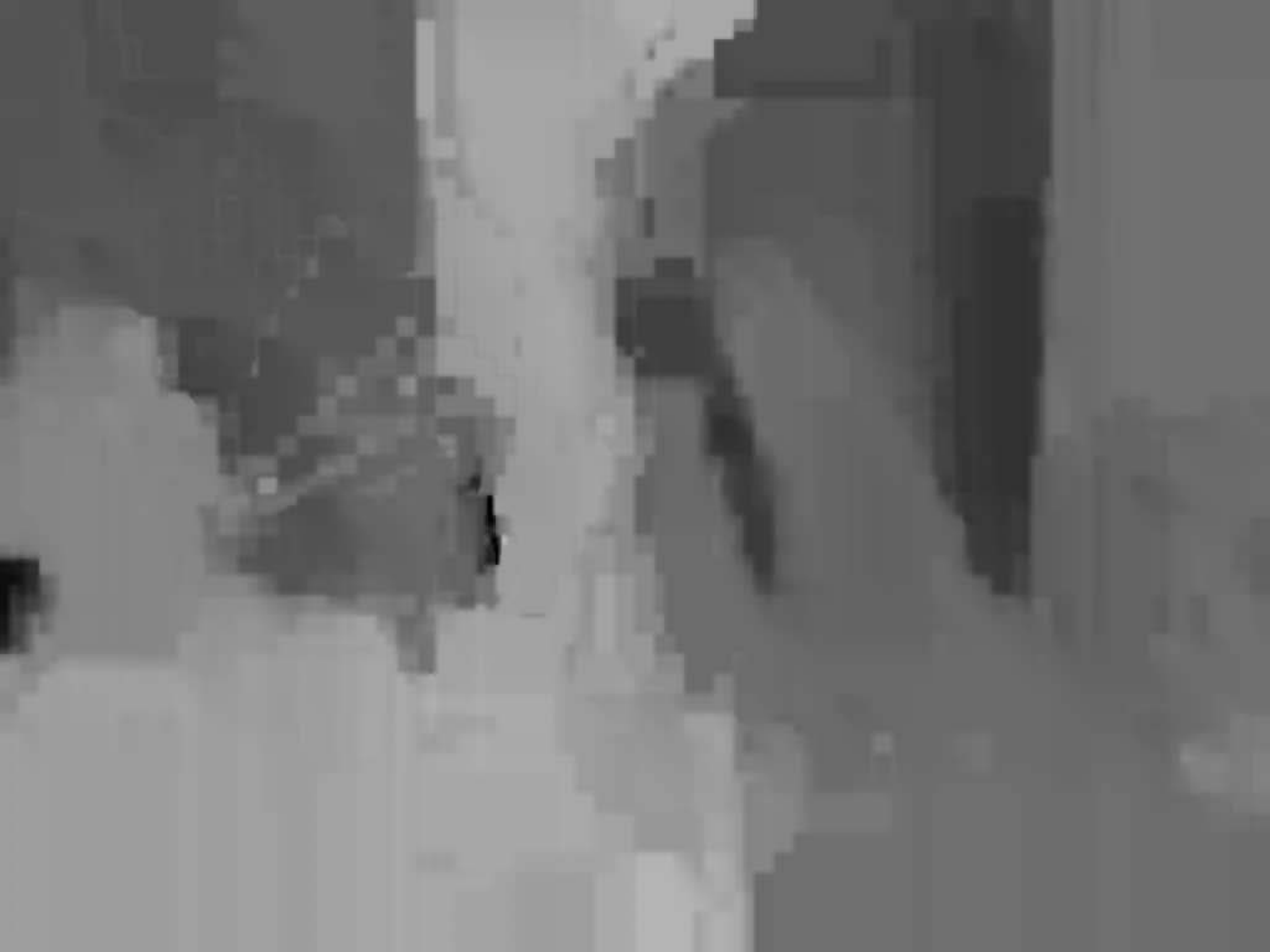}
  \centerline{\footnotesize{(e) Kendo Depth QP=49  }}
\end{minipage}

\begin{minipage}[b]{0.19\linewidth}
  \centering
\includegraphics[width=0.9\linewidth]{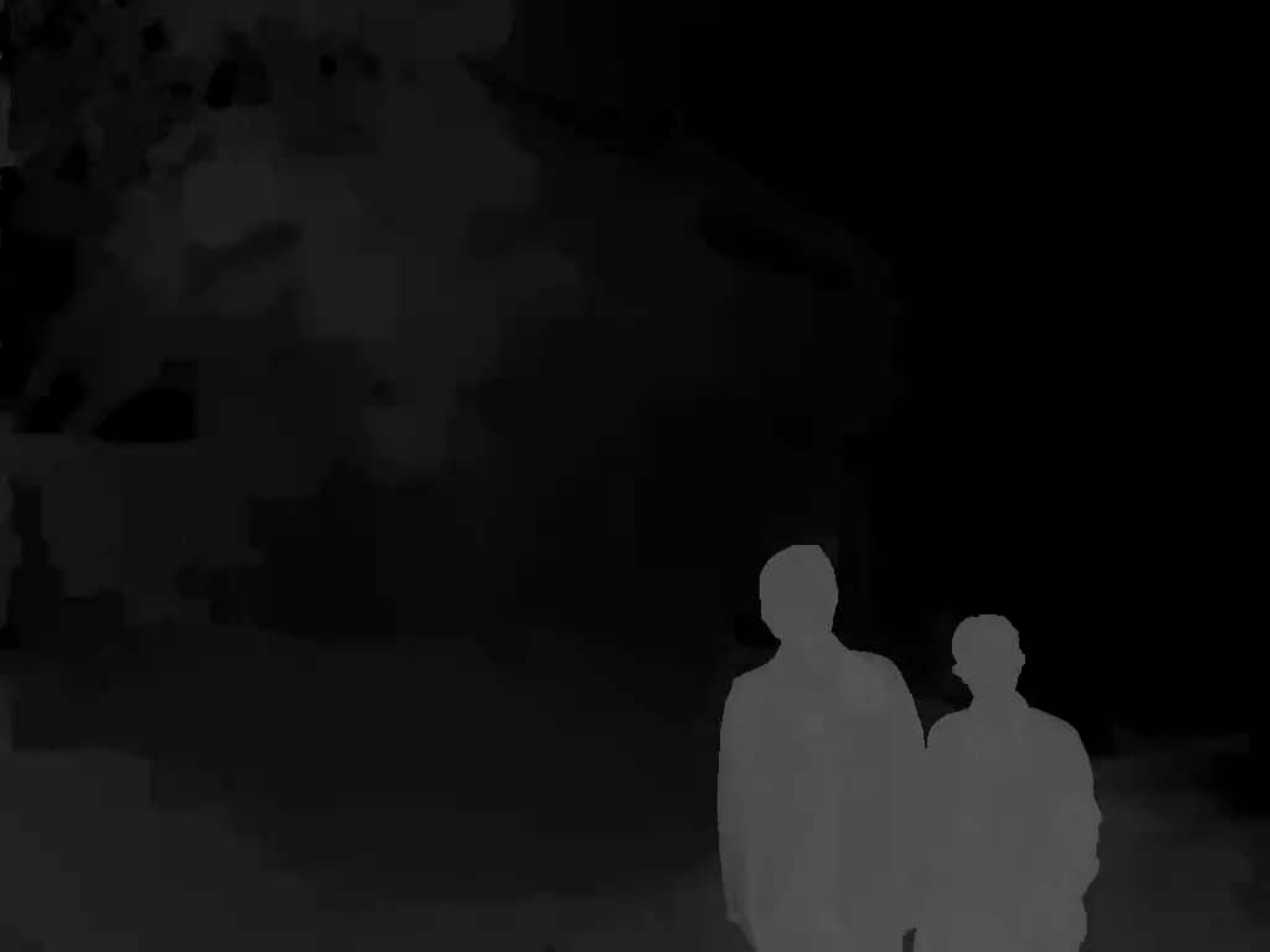}
  \centerline{\footnotesize{(f) Lovebirds Depth QP=30  }}
\end{minipage}
 \hfill
 \begin{minipage}[b]{0.19\linewidth}
  \centering
\includegraphics[width=0.9\linewidth]{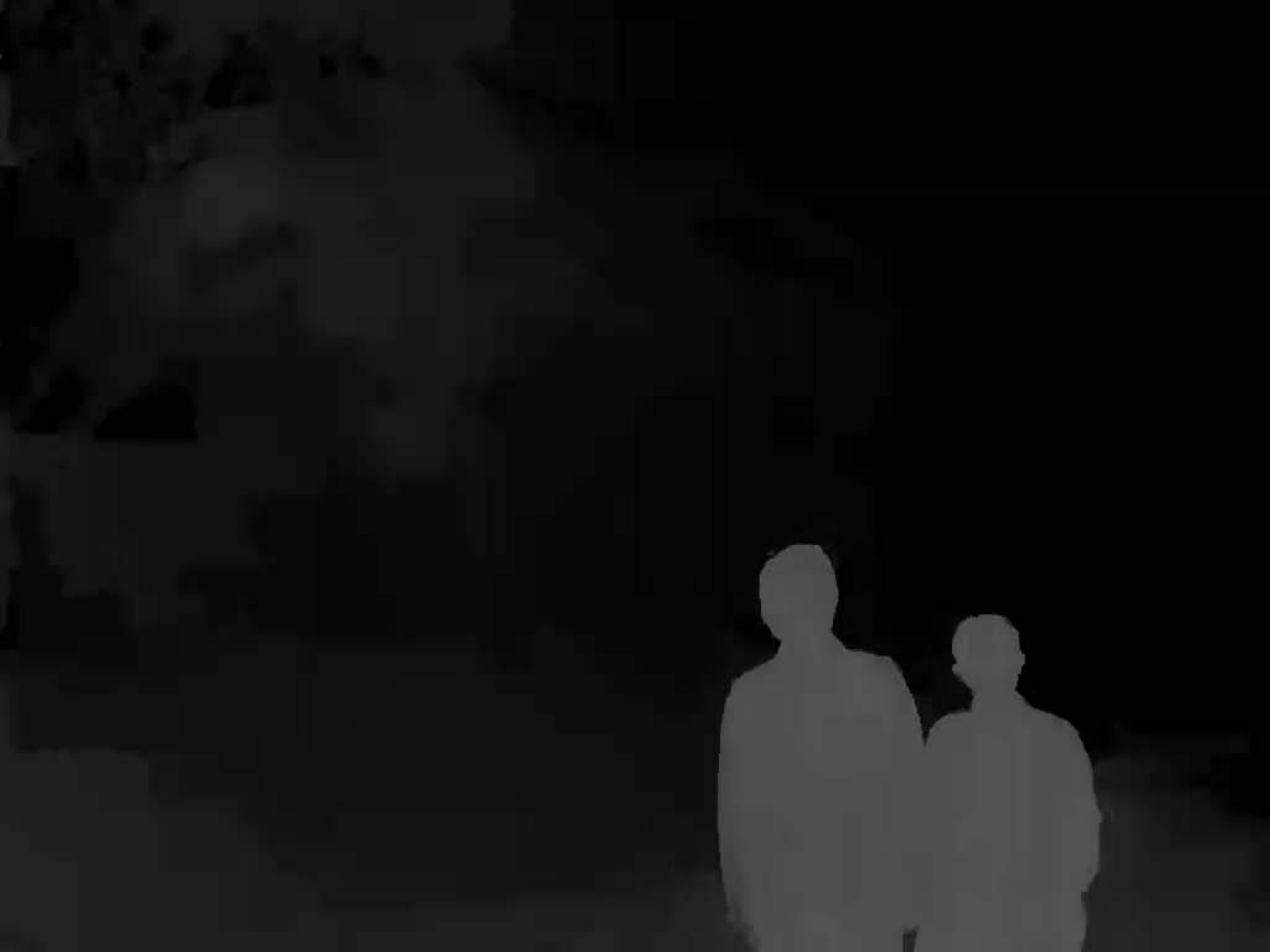}
  \centerline{\footnotesize{(g) Lovebirds Depth QP=35  }}
\end{minipage}
 \hfill
 \begin{minipage}[b]{0.19\linewidth}
  \centering
\includegraphics[width=0.9\linewidth]{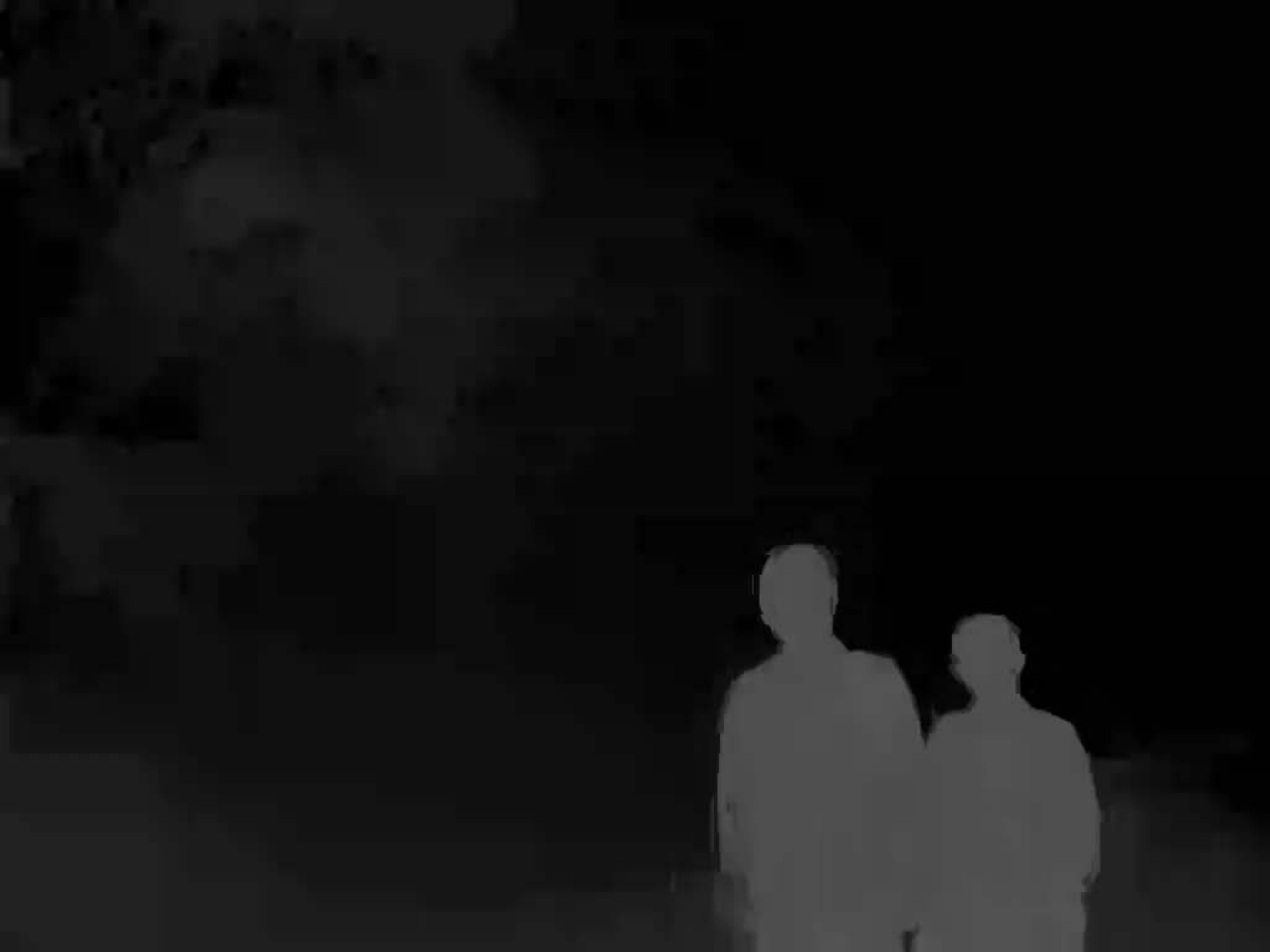}
  \centerline{\footnotesize{(h) Lovebirds Depth QP=40  }}
 \end{minipage}
 \hfill
 \begin{minipage}[b]{0.19\linewidth}
  \centering
\includegraphics[width=0.9\linewidth]{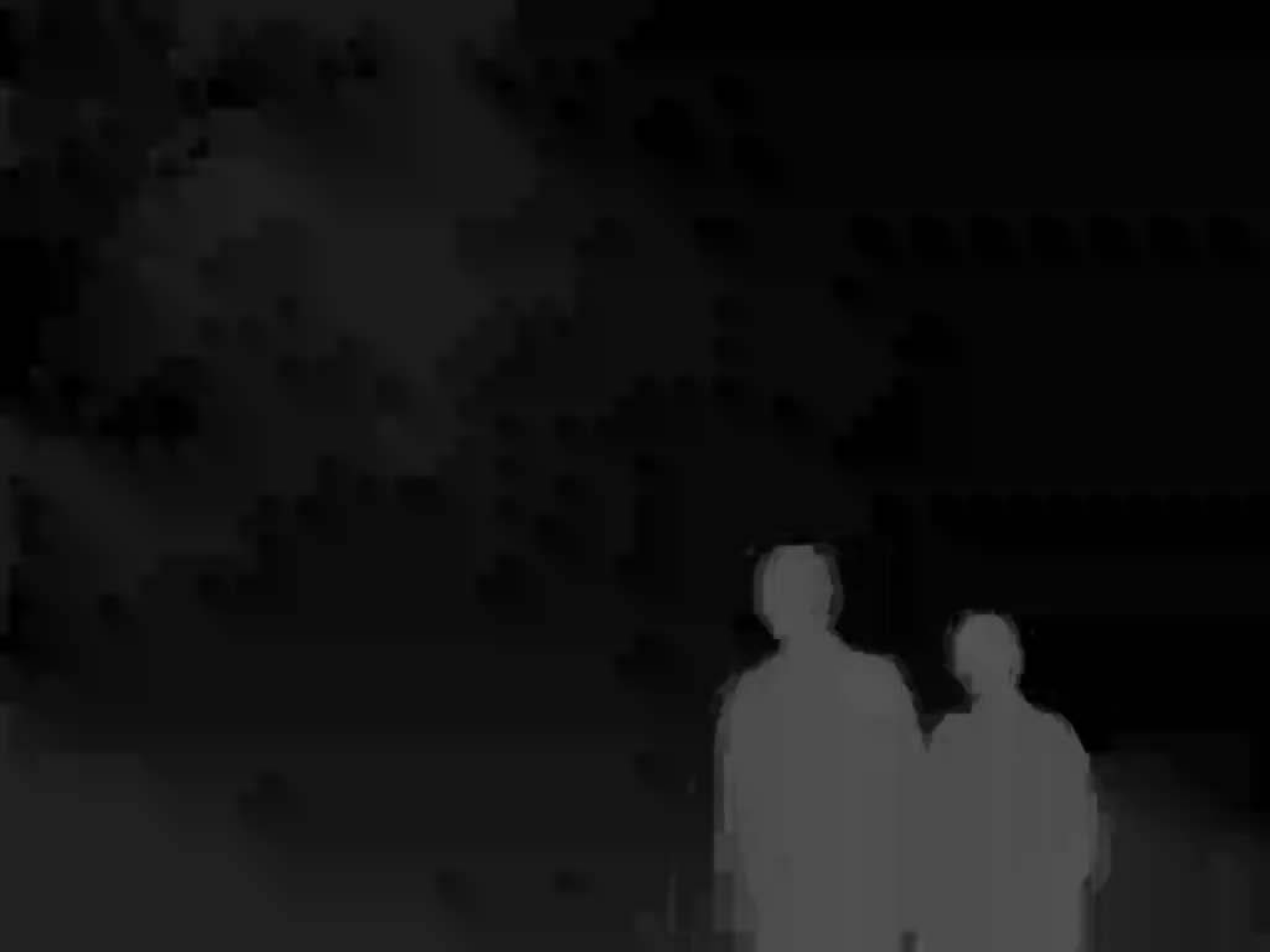}
  \centerline{\footnotesize{(i) Lovebirds Depth QP=45  }}
\end{minipage}
\hfill
 \begin{minipage}[b]{0.19\linewidth}
  \centering
\includegraphics[width=0.9\linewidth]{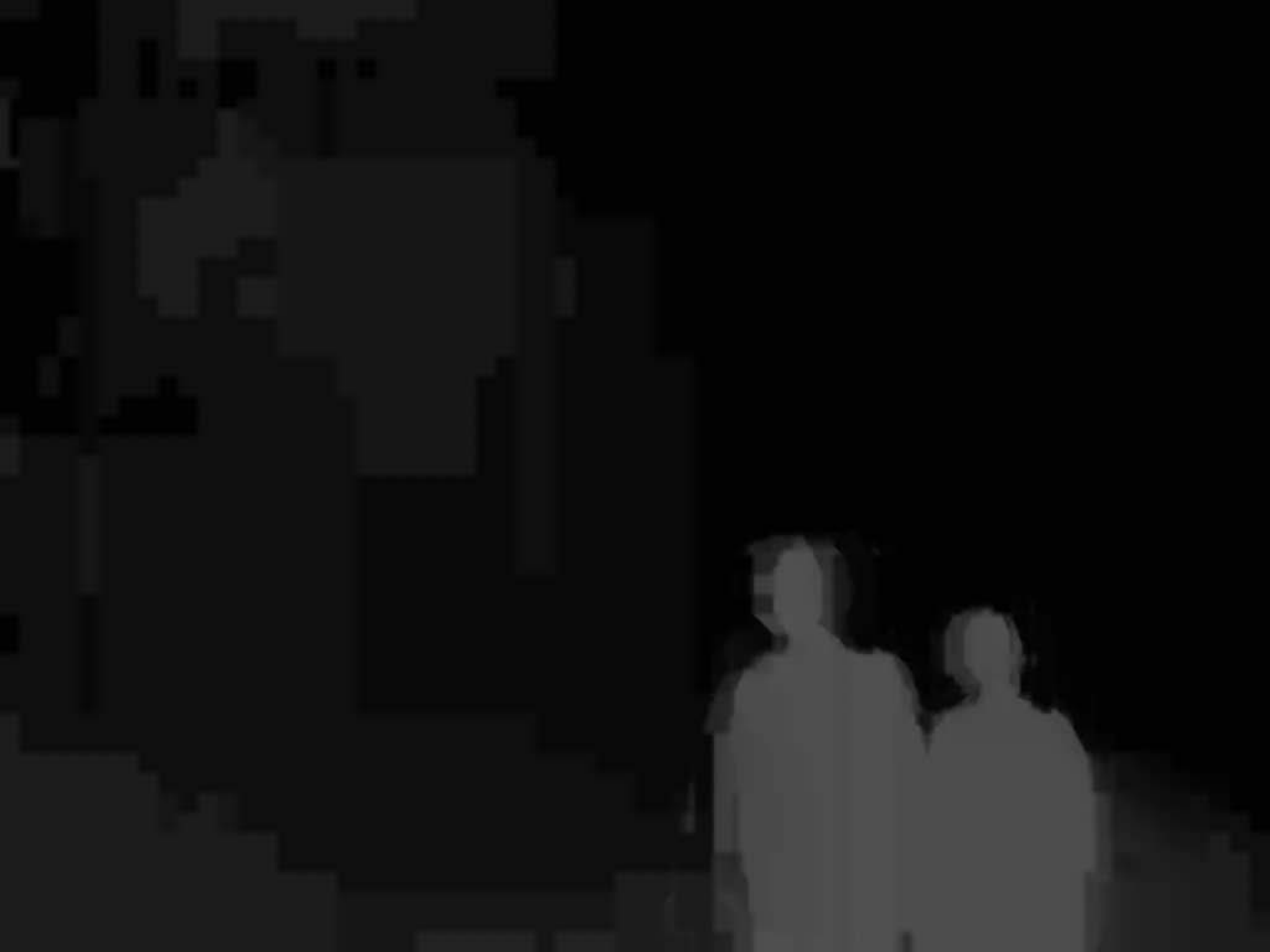}
  \centerline{\footnotesize{(j) Lovebirds Depth QP=49  }}
\end{minipage}

\caption{Compressed \texttt{Kendo} and \texttt{Lovebirds} sequences}
\label{fig:depthmaps}
\end{figure*}

It is possible to recognize the visual degradations in the depth sequences especially when \emph{QP} is set to 45 or 49 as it can be observed for \texttt{Kendo} and \texttt{Lovebirds} sequences, see Figure \ref{fig:depthmaps}. Depth maps are not directly viewed by subjects and the quality of these sequences may not highly correlate with the perceived 3D quality. However, structural deformations in the depth maps can still be a good indication of the perceived quality as explained in the rest of the section. For low QP values, visual degradations in the depth maps are not obvious, especially at low resolution. On the contrary, PSNR always shows linear decrease with the increase in QP as it is plotted in Figure \ref{fig:qualitymetrics}. Thus, we conclude that PSNR is not highly sensitive to the content and visual degradations in the depth sequences, which is understood in the research community. In contrast, \emph{VDM} and SSIM usually have a slightly steeper decrease in the last two stages. \emph{VDM} and SSIM have similar curves for \texttt{Balloons}, \texttt{Champagne Tower} and \texttt{Pantomime} sequences and they correlate with the visual degradations in the depth sequences. For \texttt{Kendo} and \texttt{Lovebirds} sequences, we can look at the compressed depth frames in Figure \ref{fig:depthmaps} to analyze the behavior of these metrics.

In order to understand the behavior of \emph{VDM} for \texttt{Kendo}, we need to consider the temporal and spatial information indexes that are given in Table \ref{tab:Indexes}. \texttt{Kendo} has the second highest spatial information index and highest temporal information index. Visual discomfort metrics vary between $0.0$ to $1.0$ and when we take higher powers of these metrics, we make \emph{VDM} more sensitive to visual discomforts. Therefore, \emph{VDM} has a steeper decrease for \texttt{Kendo} sequence. When QP is increased up to $45$ and then to $49$, we can observe that \texttt{kendo} stick loses its uniformity as it is shown in Figure \ref{fig:depthmaps}. When the pixels of the same object have different depth values, they will also have different disparity values. Thus, quantization errors such as the ones around kendo stick will cause perceivable visual discomfort that will degrade the quality of experience for the end user. The slope of the SSIM curve slightly decreases when QP is higher than $40$ which means SSIM is not highly sensitive to degradations around foreground objects that lead to visual discomfort.

\texttt{Lovebirds} depth sequence has the lowest spatial and temporal complexity as in Table \ref{tab:Indexes}. Therefore, \emph{VDM} is expected to be less sensitive to the visual discomfort. As QP is increased, we can see the blurring artifacts around foreground subjects in Figure \ref{fig:depthmaps}. If we consider the background of the depth frames more carefully, we can observe  blockiness artifacts especially when $QP=49$. These kind of blockiness artifacts degrade the quality of experience for the end user. Objective metrics are supposed to slightly decrease until QP is set to 45 and they should significantly  decrease at $QP=49$. When PSNR curve is considered, we can see that it linearly decreases as QP is increased and the slope decreases between $45$ and $49$ which contradicts with the visual degradations. SSIM significantly decreases for most of the QP values whereas it increases when QP is changed from $45$ to $49$ which negatively correlates with the visual degradations. In the case of \emph{VDM}, it highly correlates with visual degradations by slightly decreasing until $QP=45$ and  significantly decreasing when QP reaches $49$. \emph{VDM} is less sensitive to the visual discomfort compared to other sequences. However \emph{VDM} is still capable of estimating the degradations in the quality of experiences, especially for high QPs in Lovebirds sequence.

\emph{VDM} is able to estimate the visual discomfort in all of the sequences. However is can be oversensitive if the content is both spatially and temporally complex as we observe in the \texttt{Kendo} sequence, especially for the QP values between $35$ and $45$. SSIM correlates with the expected quality of experience for Balloons, Champagne Tower and Pantomime sequences. However, it is not highly sensitive to degradations in \texttt{Kendo} sequence for high QPs. In lovebirds, SSIM is oversensitive to the degradations for low QPs and it is insensitive to the expected degradations for high QPs. PSNR decreases linearly for all of the sequences and it does not highly correlate with the expected quality of experience.

\begin{table}[ht!]
  \centering
  \footnotesize

    \begin{tabular}{|c||c|c|}

\hline

   \multirow{2}[1]{*}{ \textbf{Depth Sequences}}  & \multirow{2}[1]{*}{\textbf{TInf Index}} &\multirow{2}[1]{*}{\textbf{SInf Index}} \\
    & & \\ \hline
     \textbf{Balloons}  &1.44 &2.23 \\ \hline
     \textbf{Champagne Tower}  &1.38 &2.11 \\ \hline
     \textbf{Kendo}  &1.88 &2.20 \\ \hline
     \textbf{Lovebirds}  &1.23 &1.99 \\ \hline
     \textbf{Pantomime} &1.64 &1.99  \\ \hline
    \end{tabular}%
  \caption{Spatial and Temporal Information Indexes of Depth Sequences}\vspace{-.3cm}\label{tab:Indexes}
\vspace{-0.5cm}
\end{table}

\section{RATE-DISTORTION EVALUATION}\label{sec:ratedisanalysis}

 In this paper, we compare three configurations for rate-distortion optimization. At first, we encode the depth sequences without enabling rate-distortion optimization (\emph{WRDO}) and setting quantization parameter to $40$. Secondly, we enable standard rate-distortion optimization (\emph{SRDO}) with initial $QP=40$, minimum $QP=30$ and maximum $QP=50$. The distortion metric used in default \emph{RDO} is Mean Absolute Difference (\emph{MAD}), which is a fidelity metric based on pixel-wise differences. In the final case, we perform \emph{VDM} based rate-distortion optimization (\emph{VDM-RDO}), which is explained in Section \ref{sec:ratedistopt}. Bit-rate is given in terms of \emph{kbits/sec} and fidelity is calculated with PSNR(dB) and SSIM. Rate-distortion optimization results  are summarized in Table \ref{tab:Stats}.

The standard rate-distortion optimization (SRDO) enables bit-rate savings for all sequences except \texttt{Lovebirds}. Whereas, \emph{VDM} based rate-distortion optimization (\emph{VDM}-RDO) results in bit-rate savings for all of the sequences. In terms of bit-rate, \emph{VDM}-RDO outperforms SRDO in all the sequences except \texttt{Kendo}. As mentioned in section \ref{sec:discomfortassesment}, \texttt{Kendo} has the second highest spatial information index and highest temporal information index which means \emph{VDM} is oversensitive for \texttt{Kendo}  compared to other sequences. Thus, \emph{VDM}-RDO allocates more bits to \texttt{Kendo}  to avoid visual discomfort. On average, $82.69$ $kbits/sec$ is required for WRDO, SRDO results in $54.10$ $kbits/sec$ and VMD-RDO in $38.99$ $kbits/sec$. As a trade off, rate-distortion optimization leads to lower fidelity metric values. PSNR decreases by $1.77$ dB for SRDO and $4.03$ dB for \emph{VDM}-RDO. In terms of SSIM, it remains at $0.95$ for SRDO and decreases to $0.91$ for \emph{VDM}-RDO. The main decrease in SSIM occurs at \texttt{Lovebirds} sequence for which SSIM does not highly correlate with visual degradations in depth sequences. PSNR decrease illustrates pixel-wise fidelity degradation and it does not represent perceived quality.


\begin{table}[ht!]
  \centering
  \footnotesize

    \begin{tabular}{c||c|c|c}
%

\hline


\textbf{Depth Sequences } &\textbf{WRDO} &\textbf{SRDO} &\textbf{VDM-RDO}   \\ \hline



\multicolumn{4}{c}{\textbf{Bit-rate (kbits/sec)}}     \\ \hline

    \textbf{\texttt{Balloons}}  &93.58 & 48.79 & 42.53 \\ 
    \textbf{\texttt{Champagne}} \textbf{\texttt{Tower}}  & 56.58 & 50.45 & 26.45  \\ 
    \textbf{\texttt{Kendo}}  & 153.84 & 57.83 & 67.76\\ 
    \textbf{\texttt{Lovebirds}} & 37.66 & 49.55 & 19.47 \\ 
    \textbf{\texttt{Pantomime}} & 71.81 & 48.86 & 38.75\\ 
    \textbf{\texttt{AVERAGE}} & \textbf{82.69} & \textbf{54.10}& \textbf{38.99} \\ \hline

\multicolumn{4}{c}{\textbf{PSNR (dB)}}     \\ \hline

    \textbf{\texttt{Balloons}} & 36.08 & 33.51 & 32.49 \\ 
    \textbf{\texttt{Champagne}} \textbf{\texttt{Tower}}  & 39.73 & 38.78 & 35.45  \\ 
    \textbf{\texttt{Kendo}}  & 35.11 & 30.63 & 30.96\\ 
    \textbf{\texttt{Lovebirds}} & 39.56 & 41.40 & 34.92 \\ 
    \textbf{\texttt{Pantomime}} & 39.37 & 36.70 & 35.90\\ 
    \textbf{\texttt{AVERAGE}} & \textbf{37.97} & \textbf{36.20} & \textbf{33.94}\\ \hline
\multicolumn{4}{c}{\textbf{SSIM}}     \\ \hline
    \textbf{\texttt{Balloons}} & 0.96 & 0.94 & 0.93 \\ 
    \textbf{\texttt{Champagne}} \textbf{\texttt{Tower}}  & 0.98 & 0.98 & 0.95  \\ 
    \textbf{\texttt{Kendo}}  &0.95 & 0.89 & 0.90\\ 
    \textbf{\texttt{Lovebirds}} & 0.90 & 0.97 & 0.81 \\ 
    \textbf{\texttt{Pantomime}} & 0.97 & 0.95 & 0.95\\ 
    \textbf{\texttt{AVERAGE}} & \textbf{0.95} & \textbf{0.95} & \textbf{0.91}\\ \hline
    \end{tabular}%
  \caption{Rate distortion metrics calculated over 200 frames}\vspace{-.3cm}\label{tab:Stats}
\vspace{-0.8cm}
\end{table}

\section{CONCLUSION}\label{sec:conc}

In this paper, we propose a rate-distortion optimization method for DIBR-based 3D videos. Instead of using  fidelity metrics such as PSNR and SSIM, we use content adaptive visual discomfort measure \emph{VDM}. Compared to standard rate-distortion optimization, on average, we can save $15.11$ $kbits/sec$ on bit-rate. As a price of bit-rate savings, \emph{VDM} results in 2.26 dB decrease in PSNR and 0.04 is SSIM in terms of image fidelity. The main contribution of the proposed approach is saving from the bit-rate while maintaining the quality of experience level by taking perception into consideration.





\end{document}